# Programmable Cellular Automata Based Efficient Parallel AES Encryption Algorithm

Debasis Das[1], Rajiv Misra[2]

Department of Computer Science and Engineering,
Indian Institute of Technology , Patna
Patna-800013, Bihar , India.
{ddas,rajivm}@iitp.ac.in

## ABSTRACT

*Cellular Automata(CA) is a discrete computing model which provides simple, flexible and efficient platform for simulating complicated systems and performing complex computation based on the neighborhoods information. CA consists of two components 1) a set of cells and 2) a set of rules . Programmable Cellular Automata(PCA) employs some control signals on a Cellular Automata(CA) structure. Programmable Cellular Automata were successfully applied for simulation of biological systems, physical systems and recently to design parallel and distributed algorithms for solving task density and synchronization problems. In this paper PCA is applied to develop cryptography algorithms. This paper deals with the cryptography for a parallel AES encryption algorithm based on programmable cellular automata. This proposed algorithm based on symmetric key systems.*

## 1. INTRODUCTION

A Cellular Automaton (CA)[1] is a computing model of complex system using simple rule. Researchers, scientists and practitioners from different fields have exploited the CA paradigm of local information, decentralized control and universal computation for modeling different applications. Wolfram [1] has investigated cellular automata using empirical observations and simulations. For 2-state 3-neighborhood CA, the evolution of the *ith* cell can be represented as a function of the present states of $(i-1)th$, $(i)th$, and $(i+1)th$ cells(shown in Figure 1) as: $x_i(t+1) = f(x_{i-1}(t), x_i(t), x_{i+1}(t))$ where *f*, represents the combinational logic. For a 2-state 3-neighborhood cellular automaton there are $2^3$ =8distinct neighborhood configurations and $2^8$=256 distinct mappings from all these neighborhood configurations to the next state, each mapping representing a CA rule.

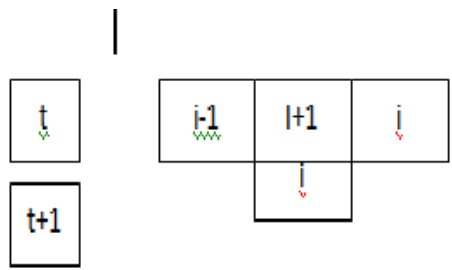

**Figure 1** : One dimentional Cellular Automata

The main aspect of cryptography and network security due to rapid development of information technology application. Cryptographic technique[2] based on two categories (1)symmetric key and (2)public key. CA based public cipher was proposed by guan[3].Stream CA based encryption algorithm was first proposed by wolfram[4]. Block encryption using hybrid additive cellular automata was proposed by Petre Anghelescu et. al[5].Cellular Automata computations and secret key cryptography was proposed by F. Seredynski et. al[6]. Block cipher based on reversible cellular automata was proposed by M. Seredynski and P. Bouvary[7].

## 1.1. Concept of Cellular Automata

Cellular Automata(CA)[1] is a collection of cells and each cell change in states by following a local rule that depends on the environment of the cell. The environment of a cell is usually taken to be a small number of neighboring cells. Figure 2 shows two typical neighborhood options (a) Von Neumann Neighborhood (b) Moore Neighborhood.

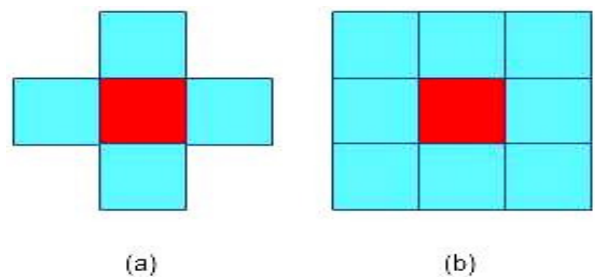

**Figure 2** : (a) Von Neumann Neighborhood (b)Moore Neighborhood

## 1.2. Concept of Programmable Cellular Automata

In Programmable Cellular Automata (PCA)[1], the Combinational Logic (CL) of each cell is not fixed but controlled by a number of control signals. As the matter of fact, PCA are essentially a modified CA structure. It employs some control signals on a CA structure. By specifying certain values of control signals at run time, a PCA can implement various functions dynamically in terms of different rules. A huge flexibility into this programmable structure can be introduced via control signals in CL. For an n-cell CA structure can be used for implementing $2^n$ CA configurations. In Figure 3 shows a 3-cell programmable CA structure and a PCA cell.

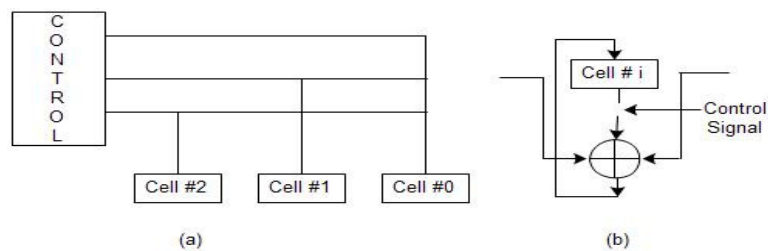

**Figure 3**: (a) A 3-cell Programmable CA Structure (b) A PCA cell

## 1.3. Type of Cellular Automata

Different variation of CA have been proposed to ease the design and modeling of complex Systems.

### 1.3.1. Linear CA

The Linear Cellular Automata have been explored by S. Nandi, B.K. Kar, and P. Pal Chaudhuri et al.[10]. If the Rule of CA involves only XOR logic then it is called the linear rules .A CA with all the cells having linear rules is called linear CA. In linear CA, the next state function applied at each cell follows the operation of Galois field(GF())[11]. The linear CA are also termed as GF(q) CA where q is a prime number.

### 1.3.2. Complement CA

The Complement Cellular Automata have been explored by S. Nandi, B.K. Kar, and P. Pal Chaudhuri et al[10]. If the Rule of CA involves only XNOR logic then it is called the Complement rules . A CA with all the cells having Complements rules is called Complement CA.

### 1.3.3. Additive CA

The Additive Cellular Automata have been explored by S.Nandi, B.K. Kar, and P. Pal Chaudhuri et al[10].A CA having a combination of XOR and XNOR rules is called Additive CA. They matrix algebraic tools that characterize Additive CA and help develop its applications in the field of VLSI testing. The Additive CA schemes based on easily testable FSM, bit-error correcting code, byte error correcting code, and characterization of 2D cellular automata. The Additive CA used in universal pattern generation, data encryption, and synthesis of easily testable combinational logic. The new characterizations of additive CA behavior , Additive CA-based tools for fault diagnosis, and a wide variety of applications to solve real-life problems.

### 1.3.4. Uniform CA

The Uniform Cellular Automata have been explored by S.Nandi, B.K. Kar, and P. Pal Chaudhuri et al[10]. If all the cells obey the same rule,then the CA said to be a Uniform CA.

### 1.3.5. Hybrid CA

The Hybrid Cellular Automata have been explored by P. Anghelescu,S. Ionita and E. Sofron et al[10].If all the cells obey the different rule, then the CA said to be a Hybrid CA. The hybrid CA has been especially applied in a linear/additive variant in which the rule set can be analyzed through matrix algebra [10]. In [11] Das has shown that a three neighborhood additive CA can be represented by a tri diagonal matrix a matrix which has the elements of its diagonal and two off-diagonals as non-zero. The properties of CA with varying (non-uniform) neighborhoods.

### 1.3.6. Null Boundary CA

The Null Boundary Cellular Automata have been explored by A. Kundu and A.R.Paul et al. [8].A CA said to be a null boundary CA if both the left and right neighbour of the leftmost and rightmost terminal cell is connected to logic 0. One-dimensional (1D) Cellular Automata (CA)over finite fields are studied in which each interior (local) cell is updated to contain the sum of the previous values of its two nearest (left & right) neighbors along with its own cell value. Boundary cells are updated according to Null Boundary conditions. For a given initial configuration, the CA evolves through state transitions to an attracting cycle which is defined as attractor / basin . The number of cycles can be determined from the minimal polynomial and characteristic polynomial of the updated matrix which is formed by the linear CA. For detailed theoretical study, follow [10]. But, in case of non-linear CA, matrix can not be formed since it does not follow any regular mathematics.

### 1.3.7. Periodic Boundary CA

The Periodic Boundary Cellular Automata have been explored by P. Anghelescu,S. Ionita and E. Sofron et al[8].In Periodic Boundary CA the rightmost cell as the left neighbour of leftmost cell. Similarly ,the leftmost cell is considered as the right neighbour of rightmost cell. So, it is like a circular linked list data structure.

### 1.3.8. Programmable CA

The Programmable Cellular Automata have been explored by P. Anghelescu,S. Ionita and E. Sofron et al[12].A CA is called Programmable CA if it employs some control signals. By specifying values of control signal at run time, programmable CA can implement various function dynamically.

### 1.3.9. Reversible CA

The Reversible Cellular Automata have been explored by M. Seredynski and P. Bouvry et al[7]. A CA is said to be reversible CA in the sense that the CA will always return to its initial state. The Interesting Property of Being the Reversible which Means that not only forward but also reverse iteration is possible. Using Reversible Rule it is always possible to return to an initial state of CA at any point. One Rule is used for forward iteration and Another Rule, reversible to the first one ,is used for backward iteration This type CA used in Cryptography.

### 1.3.10. Non-Linear CA

The Non-Linear Cellular Automata have been explored by S. Das et al[13]. In non linear CA we are used CA with all possible logic. This paper establishes the non-linear CA as a powerful pattern recognizer.

### 1.3.11. Generalized Multiple Attractor CA

The special class of *CA*, referred to as GMACA[15] *(Generalized Multiple Attractor Cellular Automata)*, is employed for the design. The desired CA model, evolved through an efficient implementation of genetic algorithm, is found to be at the edge of chaos. Cellular automata are mathematical idealizations of complex systems in discrete space and time.

### 1.3.12. Fuzzy CA:

The Fuzzy Cellular Automata have been explored by P. Maji and P. Pal Chaudhuri et al[14]. Fuzzy CA means CA with fuzzy logic. Application of fuzzy CA in pattern recognition. A special class of CA referred to as Fuzzy CA (FCA)[14] is employed to design the pattern classifier. In simple CA can handle only the Binary Patterns. In Fuzzy Cellular Automata, Each cell assumes a state and a Rational Value in [0,1].If We develop Hybrid System using CA then it is the combination of CA, Neural Network and fuzzy set or the combination of CA, Fuzzy set and Rough set.

## 1.4 . Advantages of CA in Various Research Fields

### 1.4.1. Sequential Fault Convergence

In Hardware Implementation[9] of CA, the experimental Result show that our cellular Automata produces better sequential fault convergence then the linear feedback shift register .Here we are applying the linear hybrid cellular automata rules[12].

### 1.4.2. Memorizing Capacity

The memorizing capacity of a highbred 3-neighborhood CA is better then that of Hopfield network. the Hopfield network is the model of neural network known for it association capacity.

### 1.4.3. Simulation Performance

A cellular Automata Machine can achieve simulation performance of at least several order of magnitude higher than that can be achieved with a conventional computer at compactable cost.

### 1.4.4. Theoretical Framework

A theoretical framework to study CA evolution based on graph theoretic formulation. A graph named as RVG ( Rule Vector Graph ) can be derived from the rule vector of a CA employing linear and non-linear rules. CA evolution can be characterized from the study of RVG properties.

### 1.4.5. Soft Computing

A soft computing tool for CA synthesis A methodology is under development for evolution of SOCA ( Self Organizing CA ) to realize a given global behavior.

### 1.4.6. Modeling Tools

Modeling Tools Based on the CA theory developed, a general methodology is under development to build a CA based model to simulate a system. The modeling tool enables design of a program to be executed on PCA ( Programmable CA) to simulate the given system environment.

### 1.4.7. Pattern recognition

Pattern recognition in the current Cyber Age, has got wide varieties of applications. CA based Pattern Classification / Clustering methodologies are under development based on the theoretical framework.

### 1.4.8. CA-Encompression

CA-Encompression (Encryption + Compression ) ,In the current cyber age, large volume of different classes of data - text, image, graphics, video, audio, voice, custom data files are stored and/or transferred over communication links. Compression and security of such data files are of major concern. Solutions to these problems lie in the development of high speed low cost software/hardware for data compression and data encryption. CA-Encompression technology is being developed as a single integrated operation for both compression and encryption of specific classes of data files such as medical image, voice data, video conference , DNA sequence, Protein sequence etc. Both lossy and lossless encompression are under development based on CA model.

**1.4.9. CA Compression**

Standalone CA Compression or CA-Encryption Technology Instead of a single integrated operation of compression and encryption, if a user demands only Compression or only Encryption, it can be supported using standalone packages (software / hardware version).

**1.4.10 CA Based AES**

CA based AES (Advanced Encryption System) ,As AES is the most popular security package, CA based implementation of AES algorithm in underway for development of low cost, high speed hardwired version of AES, is under development.

## 1.5. AES Encryption Algorithm

The Advance Encryption Standard [2] is a block cipher that encrypts and decrypts a data block of 128 bits. It provides extra flexibility over that required of an AES candidate, in that both the key size and the block size may be chosen to be any of 128, 192, or 256 bits but for the Advanced Encryption Standard (AES) the only length allowed is 128. It uses 10, 12 or 14 rounds[2]. The key size, which can be 128, 192 or 256 bits[2], depends on the number of round.

**1.5.1 General Design of AES Encryption**

In Figure 4 [2] shows the general design for the encryption algorithm; the decryption algorithm[2] is similar, but round keys are applied in the reverse order. In this figure-4 Nr defines the number of rounds. There is a relationship between number of rounds and the key size, which means we can have different AES versions; they are AES-128, AES-192 and AES-256. The round keys, which are created by the key-expansion algorithm, are always 128 bits, the same size as the plaintext or cipher text block.

The above figure 4 shows the structure of each round. Each round takes a state and creates another state to be used for the next transformation or the next round. The pre-round section uses only one transformation (AddRoundKey); the last round uses only three transformation(MixColumns transformation is missing).

To provide security, AES uses four types of transformations: substitution, permutation, mixing and key adding.

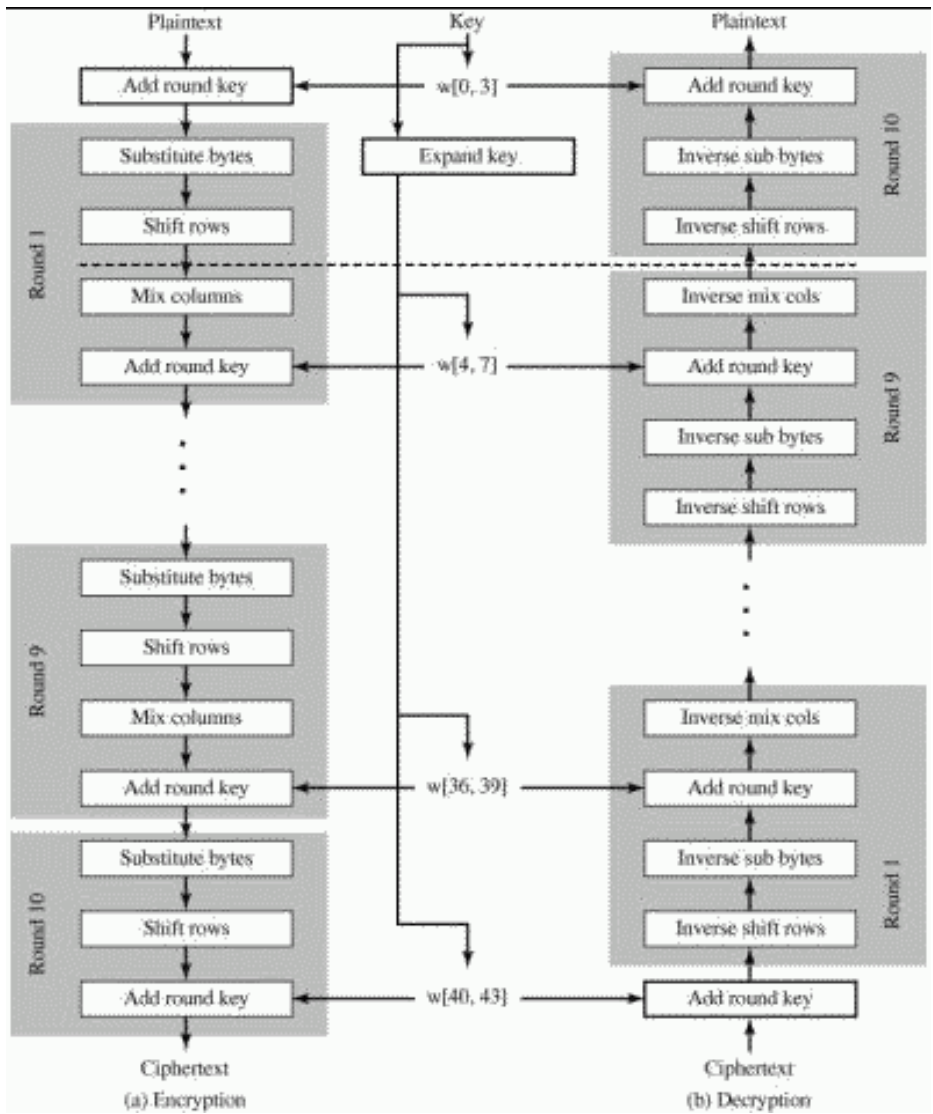

**Figure 4:** AES Block Diagram

### 1.5.1 Substitution

The first transformation, SubBytes, is used at the encryption site. In the SubByte transformation, the state is treated as a 4x4 matrix of bytes. Transformation is done one byte at a time. The SubByte operation involves 16 independent byte-to-byte transformation. This transformation is non-linear byte transformation.

InvSubByte is the inverse of SubBytes. The transformation is used at decryption site.

### 1.5.2 Permutation

Next transformation in round is shifting, which permutes the bytes. Shifting is done at the byte level. In the encryption the transformation is called ShiftRows and the shifting is to the left. The number of shifts depends on the row number(0,1,2 or 3) of the state matrix.

In the decryption, the shifting is called InvShiftRows and the shifting is to the right.

### 1.5.3 Mixing

The mixing transformation changes the contents of each byte by taking four bytes at a time and combining them to recreate four new bytes. The mixing can be provided by matrix multiplication. The MixColumn transformation operates at the column level; it transforms each column of the state to a new column. The transformation is actually a matrix multiplication of a state column by a constant square matrix.

The InvMixColumn transformation is basically the same as the MixColumns transformation and it is used at the decryption site.

### 1.5.4 Key Adding

AddRoundKey also proceeds one column at a time. AddRoundKey adds a round key word with each state column matrix.

### 1.5.5. Analysis of AES

a. AES is more secure than DES due to the larger key size. For DES we need $2^{56}$ tests to find the keys; for AES we need $2^{128}$ tests to find the key.
b. The strong diffusion and confusion provided by the different transformation removes any frequency pattern in the plaintext.
c. The algorithms used in AES are so simple that they can be easily implemented using cheap processors and a minimum amount of memory.

## 2. PROPOSED AES ENCRYPTION ALGORITHM BASED ON PCA

### 2.1 Introduction

The Programmable Cellular Automata based on the elementary CA. proposed scheme is based on two CA one is elementary CA and the other is PCA. This PCA is used to provide real time keys for the block cipher. The block diagram of programmable cellular automata encryption systems is presented in Figure 5.

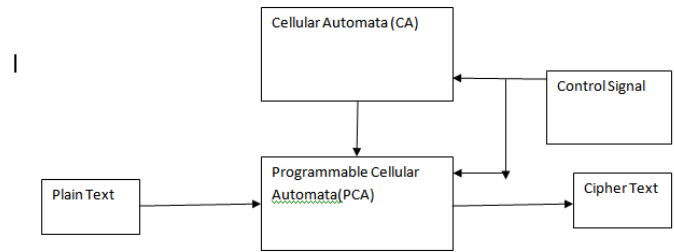

**Figure 5**: Block Diagram of AES Encryption System Based on PCA

## 2.2 Proposed Algorithm

**Algorithm: AES Enciphering and Deciphering Process Based on PCA**

**Input :** Given Plain Text / Cipher Text

**Output :** Cipher Text / Plain Text

**1 :** Enter the initial state of PCA, Convert decimal value to binary and store in an Array, A[ ],

**2:** for j=1 to $2^n$

**3 :** for i=1 to n

**4:** Apply the corresponding rule on the ith Cell, A[i].

**5:** Store the next state value, convert binary to decimal value

End of loop2 ,

End of loop1.

**6**. Create state transition diagram(or Rule Vector Graph(RVG)[8]: A Graph based on rule vector of PCA is called Rule Vector Graph. A node in RVG represents a set of RMTs(Rule Mean Time) while an edge between a pair of nodes represents the next state value (0 / 1) of a cell for specific RMTs. ) of cycle length using Rule Vector ( Rule Vector: The Sequence of rules< $R_0$, $R_1$,…$R_i$…,$R_{n-1}$> ,where ith cell is configure with rule $R_i$ ) and apply the corresponding rule.

**7 :** Insert the value of plain text into original state of PCA.

**8 :** If it is goes to its intermediate state after four cycles then

**9:** Plain Text is enciphered into cipher text.

**10 :** Else after running another four cycle the intermediate state return back to its original state.

**11:** The cipher text is deciphered into plain text

## 2.3. Rules for PCA

The rules specify the evolution of the PCA from the neighborhood configuration to the next state and these are presented in Table 1. The corresponding combinational logic of rule 51, rule 195 and rule 153 for CA can be expressed as follows:

**Rule 51: $a_i(t+1) : NOT(a_i(t))$**

**Rule 195 : $a_i(t+1) : a_{i-1}(t)\ XNOR\ a_i(t)$**

**Rule 153 : $a_i(t+1) : a_i(t)\ XNOR\ a_{i+1}(t)$**

*Table 1: The rules That Updated The next state of the CA cells :*

| Rule | 111 | 110 | 101 | 100 | 011 | 010 | 001 | 000 |
|------|-----|-----|-----|-----|-----|-----|-----|-----|
| 153  | 1   | 0   | 0   | 1   | 1   | 0   | 0   | 1   |
| 195  | 1   | 1   | 0   | 0   | 0   | 0   | 1   | 1   |
| 51   | 0   | 0   | 1   | 1   | 0   | 0   | 1   | 1   |

The operation of the simple PCA can be represented by the state transition graph. Each node of the transition graph represents one of the possible states of the PCA. The directed edges of the graph correspond to a single time step transition of the automata.

## 2.4 Procedure to Construct Transition Diagram

Considering the rule vector < 51,51,195,153> with length 4 so, the total number of states are $2^4$ = 16 states means 0000 to 1111. By using the rule vector if the start state is 0000 then next state is 1111 as shown in Figure 6 and continuing the process finally it returns back to state 0000 by completing a cycle. Initial state at time (t) : 0 0 0 0(left and right most cell connected to logic 0).

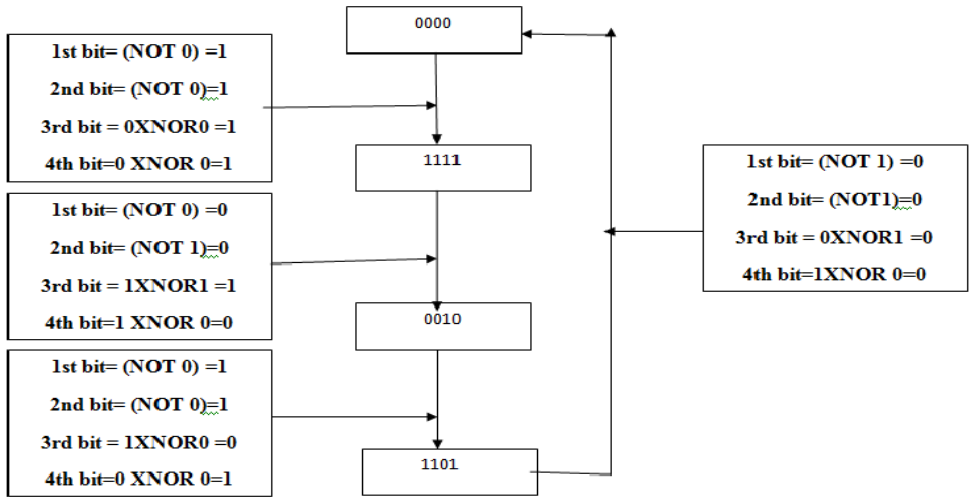

**Figure 6 :** State Changes from 0—15—2—13—0 using Rule Vector <51, 51,195, 153 >

International Journal of Network Security & Its Applications (IJNSA), Vol.3, No.6, November 2011

If the start is 0001 then next state will be 1110 (shown in Figure 7) and continuing the process finally it returns back to state 0001 by completing a cycle. Initial state at time (t) : 0 0 0 1(left and right most cell connected to logic 0).

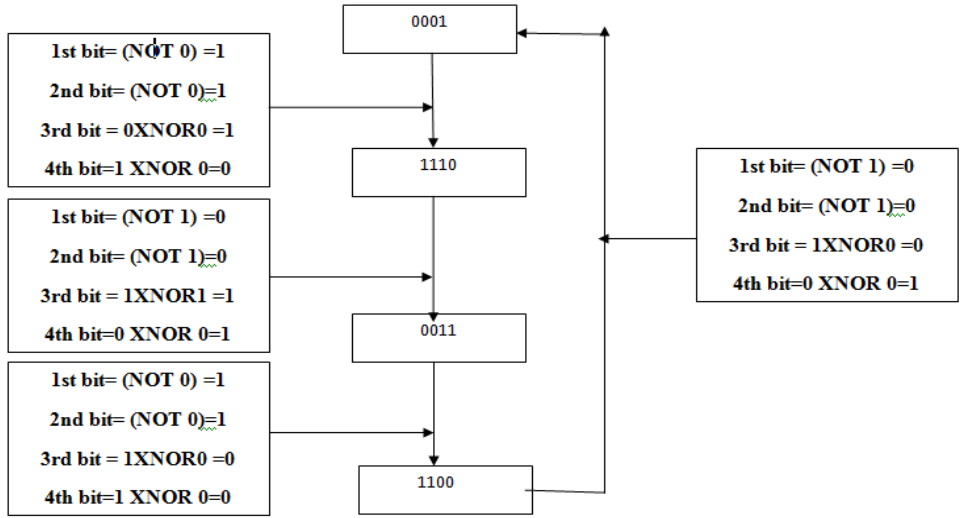

**Figure 7:** State Changes, 1—14—3—12—1 using Rule Vector <51, 51, 195, 153>

If the start is 0100 then next state will be 1001 (shown in Figure 8) and continuing the process finally it returns back to state 0100 by completing a cycle. Initial state at time (t) : 0 1 0 0 (left and right most cell connected to logic 0).

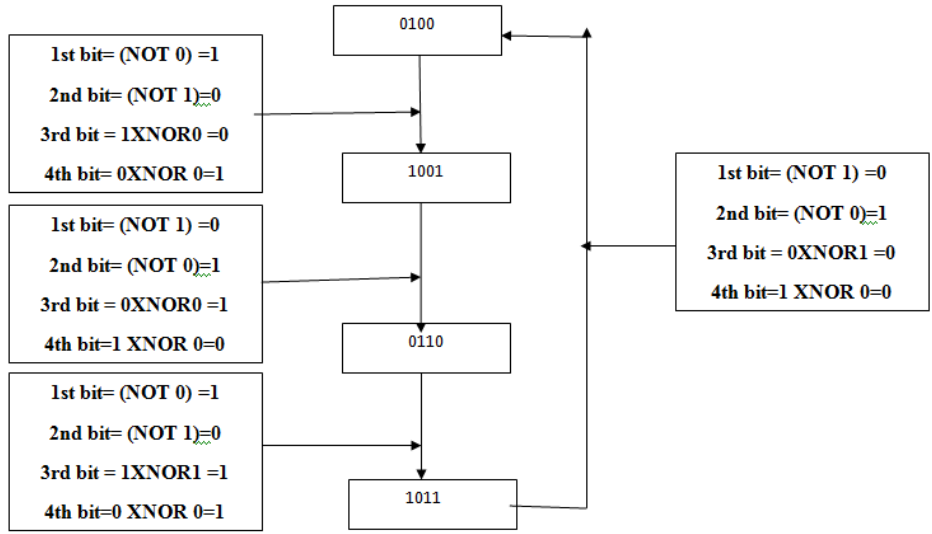

**Figure 8:** State Changes, 4—9—6—11—4 using Rule Vector <51, 51, 195, 153>

If the start is 0101 then next state will be 1000 (shown in Figure 9) and continuing the process finally it returns back to state 0101 by completing a cycle. Initial state at time (t) : 0 1 0 1 (left and right most cell connected to logic 0).

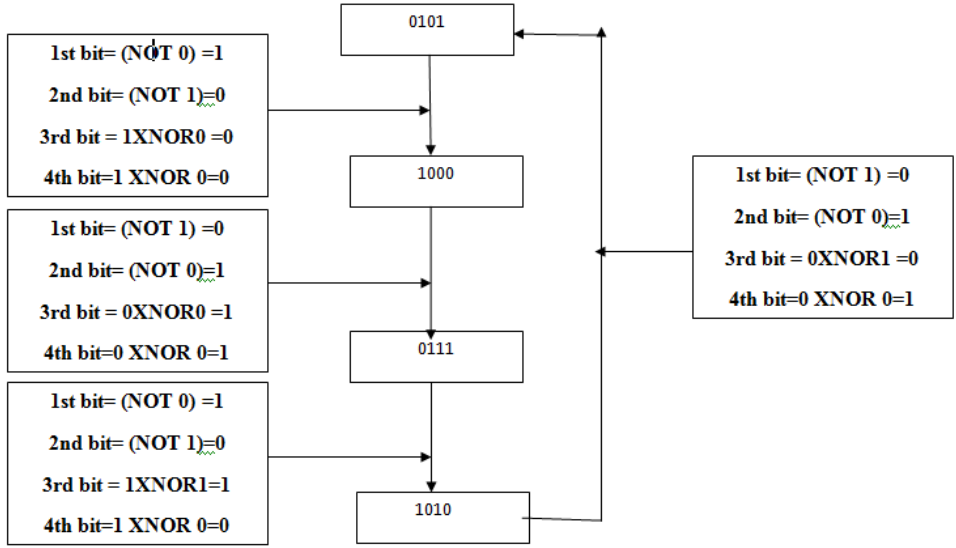

**Figure 9:** State Changes, 5—8—7—10—5 using Rule Vector <51, 51, 195, 153>

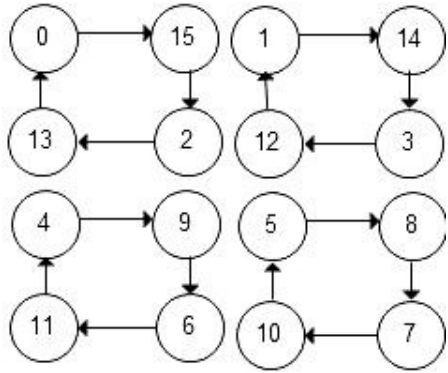

**Figure 10**: State Transition Diagram of PCA

*Table 2: Rule Selection Table*

| C1 | C2 | Rule Applied |
|----|----|--------------|
| 0  | 0  | 51           |
| 0  | 1  | 51           |
| 1  | 0  | 195          |
| 1  | 1  | 153          |

In Figure 10. the State Transition Diagram of PCA has four equal length cycles, each cycle has a cycle length 4. The rule selection table presented in Table 2. Considering this PCA as an enciphering function and defining a plain text as its original state it goes to its intermediate state after two cycles which is enciphering process. After running another four cycles, the intermediate state returns back to its original state which deciphers cipher text into plain text ensuring deciphering process.

## 3. PERFORMANCE ANALYSIS

The ICEBERG [9] scheme that proposed with the objective for efficient hardware implementation was not efficient for software implementation. The execution speed of AES code and the proposed code on a Intel Core 2 Duo 2.0 GHZ, in openMP platform. The results are tabulated in Table 3.

*Table 3: Execution Time for AES and Proposed Scheme*

| Key Size | AES           | Proposed Scheme |
|----------|---------------|-----------------|
| 128 bit  | 1.33 micro sec | 1.05 micro sec  |
| 192 bit  | 1.57 micro sec | 1.24 micro sec  |
| 256 bit  | 1.79 micro sec | 1.44 micro sec  |

Implementation speed of our scheme was found to be faster than AES for all key sizes. This could be possible due to the inherited parallelism feature of PCA. Performance result of AES and Proposed Scheme shown in figure 11. The comparision result of AES and proposed scheme based on execution time(In micro second) and different key size(128 bit, 192 bit, 256 bit).

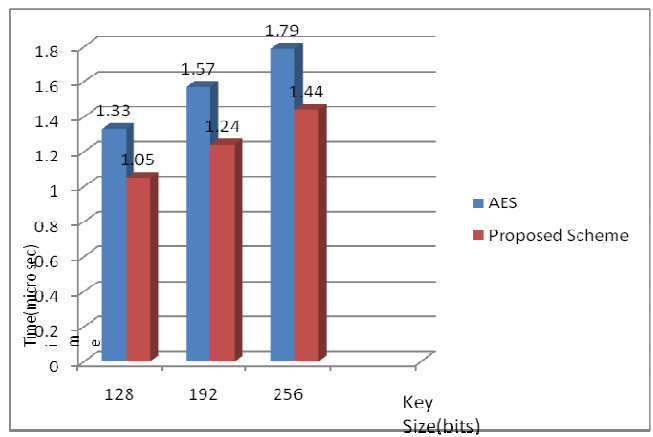

**Figure 11:** Comparision result of AES and Proposed Scheme

## 4. CONCLUSION

The proposed model in this paper presents a parallel AES encryption algorithm which is based on Programmable Cellular Automata(PCA). PCA provides higher parallelism and simplification of software implementation. The AES Encryption algorithm is being implemented on a parallel platform (OpenMP) which ensures high encryption/decryption speed. The proposed model of this paper can be implemented on other parallel platform (other than OpenMP) which ensure more security with minimum processing time. Further development of a parallel AES encryption algorithm using two CA concepts PCA and Reversible Cellular Automata (RCA). In the PCA based efficient parallel encryption algorithm , the same cipher text may be generated from different plain text which is based on the different PCA rule configuration.

## REFERENCES

[1]  S. Wolfram, ( 2002) "A new kind of science", Wolfram Media.

[2]  W. Stallings,(2003) "Cryptography and Network Security", 3rd  edition, Prentice Hall.

[3]  Guan P,(1987) " Cellular Automaton Public Key Cryptosystem",  complex system 1, pp.51- 56.

[4]  S.  Wolfram,(1985) "Cryptography with cellular Automata", pp.429-432. Springer.

[5]    Petre Anghelescu, Silviu Ionita & Emil Safron(2007) "Block Encryption Using Hybrid Additive Cellular  Automata",  7th International conference on Hybrid Intelligent Systems, IEEE.

[6]  F. Seredynski, P. Bouvry & Albert Y. Zomaya(2004), "Cellular Automata Computations and Secret Key Cryptography". Elsevier.

[7] M. Seredynski & P. Bouvary(2005) "Block Cipher Based On Reversible Cellular Automata",  New Generation Computing, pp. 245-258.,Ohmsha Ltd and Springer.

[8]   A. Kundu, A.R. Pal,  T. Sarkar,  M. Banarjee,  S. K.. Guha & D. Mukhopadhayay,(2008) "Comparative Study on Null Boundary and Periodic Boundary Neighbourhood  Mulriple Attractor Cellular Automata for Classification", IEEE.

[9]. F. Standaert ,G. Piret, G.Rouvroy , J. Quisquater, & J. Legat,(2004) "ICEBERG: An involutional Cipher efficient for block encryption in reconfigurable Hardware",  LNCS 3017. pp. 279-299. Springer Verlag.

International Journal of Network Security & Its Applications (IJNSA), Vol.3, No.6, November 2011

[10] S.Nandi, B.K. Kar & P. Pal Chaudhuri, (1994) "Theory and Applications of Cellular Automata in Cryptography" , IEEE transactions on computers, vol. 43, no. 12.

[11] A. K. Das, A. Sanyal & P. Pal Chaudhuri,(1991) "On Characterization of Cellular Automata with Matrix Algebra", Information Science.

[12] Petre Anghelescu, Silviu Ionita & Emil Safron,(2008) "FPGA Implementation of Hybrid Additive Programmable Cellular Automata", Eight International conference on Hybrid Intelligent Systems,IEEE.

[13] S. Das,(2006) "Theory and Applications of Nonlinear Cellular Automata In VLSI Design", PhD thesis, B. E. College.

[14]P. Maji & P. Pal Chaudhuri,(2004) "A Fuzzy Cellular Automata Based Pattern Classifier", DASFAA, LNCS-2973, pp.494-505.

[15] Niloy Ganguly ,P. Maji, A. Das, B. K. Sikdar,&P. Pal Chaudhuri ,(2002) "Characterization of Non-linear Cellular Automata Model for Pattern Recognition", AFSS 2002, LNAI 2275, pp. 214–220, Springer.

## Authors:

**Mr. Debasis Das** is currently pursuing Ph.D in Computer Science and Engineering from Indian Institute of Technology Patna, India. He received M. Tech in Computer Science and Engineering degree from KIIT University, Bhubaneswar in 2010. His research interests include Computer Network, Algorithm, Network Security and Cellular Automata.

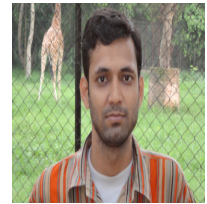

**Dr. Rajiv Misra** is currently working as Assistant Professor in Department of Computer Science and Engineering in Indian Institute of Technology Patna, India. He received Ph.D from IIT Kharagpur in field of Mobile Computing in 2010. He holds M Tech degree in Computer Science and Engineering from the Indian Institute of Technology (IIT), Bombay, in 1989and BE degree in Computer Science from the MNIT Allahabad, in 1987. His research interests include Mobile Computing, Ad hoc Networks and Sensor Networks, Vehicular Networks and Intelligent Transportation System. He has published papers in IEEE Transaction in Mobile Computing and IEEE Transaction in Parallel and Distributed Systems. He is a member of the IEEE.

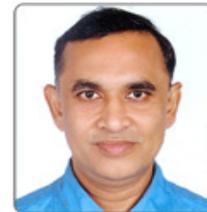